\documentclass[final,5p,times,twocolumn,authoryear]{elsarticle}

\usepackage{amssymb}
\usepackage{amsmath}
\usepackage{lipsum}
\usepackage{booktabs}
\usepackage{multirow}

\newcommand{\kms}{km\,s$^{-1}$}

\journal{New Astronomy}

\begin{document}

\begin{frontmatter}



\title{Low-frequency observations of low-mass binary systems with neutron star candidates}


\author[1]{Elena Brylyakova}
\ead{elinxt@bk.ru}
\author[2,3]{Marina Afonina}
\ead{afonina.md19@physics.msu.ru}
\author[4]{Gayane Tyul'basheva}
\ead{g.tyulbasheva@yandex.ru}
\author[2]{Sergei B. Popov\corref{cor1}}
\ead{sergepolar@gmail.com}
\author[1]{Sergei Tyul'bashev}
\ead{serg@prao.ru}

\affiliation[1]{organization={Lebedev Physical Institute of Astro Space Center of Pushchino Radio Astronomy Observatory},
            addressline={PRAO, Radiotelescopnaya 1a}, 
            city={Moscow region},
            postcode={142290},
            country={Russia}}
\affiliation[2]{organization={Sternberg Astronomical Institute, Lomonosov Moscow State University},
            addressline={Universitetskij pr. 13}, 
            city={Moscow},
            postcode={119234},
            country={Russia}}
\affiliation[3]{organization={Department of Physics, Lomonosov Moscow State University},
            addressline={1/2 Leninskie Gory}, 
            city={Moscow},
            postcode={119991},
            country={Russia}}
\cortext[cor1]{Corresponding author}

\affiliation[4]{organization={Institute of Mathematical Problems of Biology, Branch of Keldysh Institute of Applied Mathematics of Russian Academy of Sciences},
            addressline={IMPB RAS, Professor Vitkevich 1}, 
            city={Moscow region},
            postcode={142290},
            country={Russia}}

\begin{abstract}

 Recently, astrometric and spectroscopic observations resulted in the discovery of several low-mass binaries with invisible components, which are expected to be compact objects. In about two dozen cases, the masses of these components are consistent with neutron stars. We use low-frequency archival data obtained with the Large Phased Array in Pushchino to search for radio emission from five of these systems. For all the systems, we do not detect persistent or periodic emission. In one case (2MASS J1527+3536), we identify a single radio burst with a flux of 13 Jy and a duration of 0.13 s.  However, the dispersion measure of the burst does not correspond to an expected value for the source. We discuss several possibilities to explain the properties of this burst. 

\end{abstract}




\end{frontmatter}




\section{Introduction}
\label{introduction}

 Being interesting objects by themselves, neutron stars (NSs) also play an important role as a testbed for fundamental physics \citep{2022hxga.book...30N} and probes for their astrophysical environment, including interstellar medium \citep{2008ASSP...12...99H} and stellar winds \citep{2010ASPC..422...57N, 2017SSRv..212...59M}.
That is why, after the discovery of a new type of system involving NSs, it is necessary to perform multi-wavelength (and multi-messenger, if possible) observations of these sources.

 Recently, several NS candidates have been found in low-mass binary systems via astrometric \citep{2024NewAR..9801694E} and radial velocity \citep{2024ApJ...969..114L} measurements. 
 The systems discovered by {\it Gaia} are wide, with orbital periods of about a year.
 In contrast, spectroscopic systems have orbital periods typically below a day.
 These NSs do not manifest themselves as radio pulsars, accreting X-ray pulsars, thermal pulsating sources, bursting magnetars, etc.  \citep{2024NewAR..9801694E, 2024A&A...686A.299S}.
It is expected that when all {\it Gaia} data are reduced, the number of such binaries will increase by 1-2 orders of magnitude \citep{2025arXiv250821805C}. 
 
 As the secondary components of the binaries are low-mass main sequence stars, we can conclude that, on average, these NSs have ages of about a few billion years. Thus, any type of activity due to magnetic field decay, rotational energy losses, residual heat, etc., is expected to cease.   For most of the systems, the only source of matter for accretion is a weak stellar wind with an expected rate $\lesssim 10^{-14}$--$10^{-13}\, M_\odot$~yr$^{-1}$  (the systems are not tight enough to provide accretion via Roche-lobe overflow; however, the situation with the tightest systems discovered via spectral observations is more complicated as the stars can be very close to fill their Roche lobes, see e.g., \citet{2023ApJ...944L...4L}). This would correspond to the maximal luminosity $\sim 10^{32}$~erg~s$^{-1}$ in the case of accretion onto an NS. However, X-ray observations of some of these systems, as well as a comparison to X-ray catalogues, result in smaller upper limits \citep{2024A&A...686A.299S}.  

NSs are known or proposed to be radio transient sources of various kind: rotating radio transients \citep{2011BASI...39..333K, 2025MNRAS.537.1070T}, fast radio bursts \citep{2020Natur.587...45Z}, long-period radio transients \citep{2024arXiv241116606W}.  
In this paper, we present an archival search for radio bursts or continuous emission at the low frequency 111 MHz from a selected sample of binary systems with non-visible NS candidates. In the next section, we briefly describe the sample. Then we provide details on the observations and data reduction. In Sec.~4, we present our results.  We discuss possible mechanisms of transient radio emission from evolved NSs in low-mass binary systems in Sec.~5. In the final section, we summarize our conclusions.






\begin{table*}[t]
    \centering
    \renewcommand{\arraystretch}{1.4}
    \begin{tabular}{cccccl}
    \hline
         System & Distance, pc & $P_\mathrm{orb}$, d & Eccentricity & $M_*, \, M_\odot$ & Ref. \\ \hline
      J0616+2319  & 1111 & 0.8666 & $\approx 0$ & 1.7 & Sbarufatti et al. (2024)\\
      J1527+3536  & 118 &  0.2557 & $\approx 0$ & 0.62 & Sbarufatti et al. (2024)\\
      J2102+3703  & 657 &  481.04 & 0.45 & 1.03 & El-Badry et al. (2024)\\ 
      J2128+3316  & 227 & 1430    & 0.59 & 0.7 & Sbarufatti et al. (2024)\\ 
      J2145+2837  & 242 &  889.5  & 0.58 &  0.95 & El-Badry et al. (2024)\\ 
           \hline
    \end{tabular}
    \caption{Candidate low-mass binary systems with non-active NSs.  All data from \cite{2024OJAp....7E..58E}  or \cite{2024A&A...686A.299S}.
    }
    \label{tab_prop}
\end{table*}

\section{Sample}
\label{sample}

In this study, we used five sources from the samples presented by \cite{2024OJAp....7E..58E} and \cite{2024A&A...686A.299S}. 
Their basic properties are given in Table~\ref{tab_prop}. The first large group of 21 sources was identified by \cite{2024OJAp....7E..58E} using {\it Gaia} astrometric data. Systems in the second group  (six objects) were identified either by {\it Gaia} astrometry or by spectroscopic (radial velocity) measurements with {\it LAMOST},  \cite{2024A&A...686A.299S} presented X-ray observations of these sources. The two samples do not overlap. 
We present results for two sources from \cite{2024OJAp....7E..58E} and for three from \cite{2024A&A...686A.299S}.   Our sample includes both: wide (astrometric) and tight (spectral) systems. The distances to these sources are known, and therefore, it is possible to estimate the expected dispersion measure (DM) in their direction. 

The longest series of observations on Large Phased Array (LPA) (see Sec.~3), which can provide the best sensitivity, are collected for declinations of $-9^\circ < \delta < +42^\circ$. The selected sources are within the range of these declinations. In addition, we use only relatively close binary systems with an expected DM in the direction of these systems $<250$~pc~cm$^{-3}$, as for low-frequency observations, the peak flux is significantly reduced for highly dispersed signals.\footnote{No sources with DM~$>250$~pc~cm$^{-3}$ are found in the LPA blind search programs so far (https://bsa-analytics.prao.ru/pulsars/known /). At such a DM, the pulse scattering reaches 450~ms and even more, which strictly limits the periods of pulsars available for search \citep{Bhat2004,Kuzmin2007}.} 
Finally, for some sources, the interference has been too high to allow for a useful analysis. 

In Table \ref{tab_prop}, we present some basic parameters of the systems under study.  All stellar companions are at the main sequence stage of their evolution. We do not show small uncertainties in orbital periods for all the binaries. In \cite{2024OJAp....7E..58E}, the authors do not provide distances, but parallaxes, $\varpi$. Thus, we calculate distances (in parsecs) as $d=1/\varpi$ and do not provide the uncertainties.  All long-period systems have large eccentricities, whereas short-period binaries have $e\approx 0$; all these values are presented in the table. 

Four sources from our list might be NSs as their masses fall in the range appropriate for this type of object. A word of caution might be said about the source 2 MASS J1527+3536 (hereafter, J1527+3536). This source was identified as a binary with an invisible component due to spectroscopic measurements. \cite{2023ApJ...944L...4L} reported the mass of the invisible component of the binary as $0.98\pm 0.03\, M_\odot$.  However, \cite{2024ApJ...961L..48Z} give a different mass estimate: $0.69\pm 0.02\, M_\odot$. Then, it might be a white dwarf (WD). We discuss various possibilities for this source in Sec.~5.6 below.

\section{Observations and processing}
\label{obs}

The search for periodic and pulse radiation in binary systems was carried out using archived data from LPA radio telescope of the Lebedev Physics Institute (LPI). The radio telescope has two independent antenna patterns: LPA1 and LPA3. The latter is used in this study. The LPA3 radiation pattern consists of 128 uncontrolled (stationary) beams aligned in the meridian plane and covering the declinations of $-9^\circ < \delta < +55^\circ$. The transit time of the source through the meridian is approximately 3.5~min. The sizes of the radiation pattern are $0.5^\circ \times 1^\circ$, so typical coordinate errors are $10^\prime-15^\prime$ in declination and $0.5^m-1^m$ in right ascension \citep{Tyulbashev2016}.

Synchronous recording of LPA3 beams in two time-frequency modes has been conducted daily and around the clock since August 2014.  In this study, we use the data obtained in the interval from August 16, 2014, to June 15, 2025 (3956 days). The sampling of the point is 100 and 12.5~ms. The 2.5~MHz band is divided into 6 and 32 frequency channels with a width of 415 and 78~kHz. The LPA has an effective area of about 45~000 square meters, which provides high sensitivity at an observation frequency of 111 MHz \citep{Shishov2016,Tyulbashev2016}. The accumulated volume of raw data is currently about 450~TB. 

The archived data allow searching for both periodic and pulsed radiation. Periodic radiation is searched using the spectra obtained using the Fast Fourier Transform (FFT) and Fast Folding Algorithm (FFA). As early processing of LPA3 observations shows, the FFT search is limited to periods $P<2-3$~s, since low-frequency (red) noise sharply reduces sensitivity for periods $P >2-3$~s. The FFA search enables one to work with different periods, including $P>2$~s. However, the signal processing time is much longer in comparison to the FFT algorithm \citep{Tyulbashev2022,Tyulbashev2024}.

To increase sensitivity, we summarize the FFT/FFA spectra for the entire available observation interval. To reduce the number of calculations, FFA spectra are calculated only for periods $P>2$~s. The sensitivity of searching for periodic signals in the summed spectra is proportional to the square root of the number of spectra used. Taking into account the long-term series of observations, the sensitivity to periodic signals increases tenfold \citep{Tyulbashev2022}.

When searching for pulsed radiation, the data in the frequency channels are calibrated using a calibration signal with a known temperature. Then we add up the frequency channels, assuming different DMs. 
If the signal-to-noise ratio (S/N) is greater than 7, the dynamic spectrum and the resulting pulse profile are stored \citep{Tyulbashev2018}. To increase sensitivity, a pulse search is performed after averaging the data; then it is assumed that the pulse width is 1, 2, 4, or 8 points. We also search for periodicity after convolving the signal with a template, the width of which depends on the DM. The width is set according to the model considered by \cite{Kuzmin2007}. Averaging or convolution procedures allow for increasing the S/N of wide pulses or pulses that are widened due to scattering. 


 The DM estimates for the sources in our sample (see Table~2) are obtained from the known distances using the online tool \citep{Kaplan2022}\footnote{https://pulsar.cgca-hub.org/compute} for two models of the electron distribution in the Galaxy: NE2001 \citep{Cordes2002} and YMW2016 \citep{Yao2017}.


\section{Results}
\label{sec_res}

In this section, we present the results of the archival data analysis for five low-mass binaries with NS candidates. In the first subsection, the results of the periodicity search for all the sources are summarized. In Sec.~4.2, we present properties of a single burst detected in the direction of the source  J1527+3536.

\subsection{Limits on the periodic emission}

Periodic emission in the direction of all five sources -- J0616+2319, J1527+3536, J2102+3703, J2128+3316, and J2145+2837 -- was not found. The search was carried out using a specially developed program with a DM trial in the range of 0-1000 pc cm$^{-3}$ \citep{Tyulbashev2022}. To extract upper estimates of the integral flux density ($S_i$), it is necessary to take into account the background temperature in the direction of the candidates under study. It is also necessary to take into account the height of the source above the antenna and the discrepancy between the coordinates of the candidates and the direction to the center of the beam pattern. In addition, for small periods, the main limitation in obtaining estimates of $S_i$ is related to the pulse scattering. 

Table \ref{tab_lim} presents the upper limits of $S_i$ at the frequency 111 MHz in mJy. The first column of the table contains the names of the sources. The second column shows the ranges of the expected DM in the direction of the binary systems according to the models NE2001 and YMW2016 \citep{Cordes2002,Yao2017}. In columns 3-6, we present the upper limits on $S_i$ in mJy for different ranges of the period. Each $S_i$ limit is calculated basing on the worst-case sensitivity value within the specified period range. 
Details on the assessment of the sensitivity of LPA3 observations can be found in \cite{Tyulbashev2022}.

\begin{table*}[t]
    \centering
    \renewcommand{\arraystretch}{1.4}
    \begin{tabular}{cccccc}
    \hline
   System  & DM, pc~cm$^{-3}$ & $0.25<P\le 0.5$, s& $0.5<P\le 1.0$, s  & $1.0<P\le 2.0$, s  &$2<P<60$, s  \\ \hline
J0616+2319 & {41.1-50.8}   &  ${<0.48}$ &  ${<0.38}$              &${<0.32}$      & $<2.5$ \\
J1527+3536 & {0.6-1.3}   &  $<0.25$          &  $<0.25$             & $<0.25$     & $<1.5$ \\
J2102+3703 & {7.3-10.0}  &  ${<0.42}$     &  ${<0.37}$              & ${<0.15}$      & $<2.6$ \\ 
J2128+3316 & {2.2-3.1}    &  ${<0.25}$           &  ${<0.22}$             & ${<0.1}$     & $<1.6$\\ 
J2145+2837 &{2.4-3.3}     &  ${<0.28}$           &  ${<0.24}$              & ${<0.1}$     & $<1.7$\\ 
           \hline
     \end{tabular}
    \caption{Upper limits of $S_i$ at 111 MHz in mJy for different period boundaries.
    }
    \label{tab_lim}
\end{table*}

It is illustrative to scale the upper limits from Table~2
to the frequency 1.4 GHz using a power-law spectrum $S\sim\nu^{-\alpha}$. 
For $\alpha=2$, limits are $\sim159$ times smaller ($\lesssim 0.003$ mJy for the typical values in the table), and for $\alpha=1.5$ -- approximately 45 times smaller than at 111 MHz ($\lesssim 0.01$~mJy). 


\subsection{Detection of a radio pulse in the direction of J1527+3536}

\begin{figure}[t]
    \centering
        \includegraphics[width=0.9\linewidth]{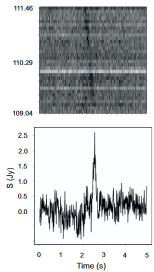}
    \caption{The dynamic spectrum of the pulse (upper panel) and the pulse profile  (lower panel) of the burst from J1527+3536. 
    In both panels, the range on the horizontal axis is equal to 5 seconds.  
    On the vertical axis of the upper panel, the frequency is given in MHz. 
    The DM~$=24.3$~pc~cm$^{-3}$ is assumed.
    }
    \label{fig_prof}
\end{figure}

When searching for dispersed pulses, a pulse was found in the direction of the source J1527+3536.
The detected pulse has S/N=8.1.  The half-width of the pulse ($W_e$) is 130~ms. The right ascension 
of the pulse is $\alpha_{2000}= 15^h31^m03^s$, which, taking into account the half-width of the beam $3.5/\cos(\delta)$ min,\footnote{Note, that the width of the beam is given in time units. } is in correspondence with the coordinate of the binary system ($\alpha_{2000}=15^h27^m48^s$). To obtain the declination of the pulse, we search for a pulse with a known DM in the beams above and below the one in which the burst has been detected. A weak signal is detected in the beam above. Knowing the shape of the radiation pattern $[\sin(x)/x]^2$ and the ratio of the signal amplitudes in the two beams, we estimate the pulse's declination as $\delta_{2000}=35^\circ35^\prime$. Taking into account the distance between the beams, which is $29^\prime$ at this declination, this value of $\delta$ is in correspondence with the declination of the system ($\delta_{2000}=+35^\circ36^\prime57^{\prime \prime}$).

After the identification of the pulse, the DM was reassessed. The result from the fit of our observations -- 
DM~$=24.3$~pc~cm$^{-3}$ -- is  not in agreement with the estimates based on the NE2001 and YMW2016 models \citep{Cordes2002,Yao2017}.  These models predict a much lower value. The detected value corresponds to the expected DM for a source in the Galactic halo, see Sec.~5.

The apparent peak flux density ($S_\mathrm{p}$) of the pulse for S/N=8.8 (at DM~$=24.3$~pc~cm$^{-3}$) corresponds to 2.7~Jy. However, in order to obtain a correct estimate of $S_\mathrm{p}$, it is necessary to take into account factors that reduce the apparent pulse flux density. Thus, the signal was detected between the two beams in declination and almost at the boundary of half the power of the radiation pattern in right ascension. This leads to large corrections for the source hitting the edge of the radiation pattern of the LPA3 beam. When passing through the meridian, the source is not at the zenith, which leads to a correction for the cosine of the source height above the antenna. The value of the correction coefficients can be seen in Fig. 1 in the paper \cite{Shishov2016}. Taking into account all the corrections, we obtain $S_\mathrm{p}=13$~Jy.

If the burst is indeed related to the system J1527+3536 (see Sec.~5), the observed isotropic radio luminosity of the pulse can be estimated as $L_\mathrm{r, iso}= S_\mathrm{p} \, 4\pi d^2 \, \Delta \nu$, where $\Delta \nu$ is the bandwidth. This is a lower limit as we have no information about the spectrum out of the relatively narrow observational band.
For $S_\mathrm{p}=13$~Jy, $\Delta \nu=2.5 \times 10^6$~Hz, and $d=118$~pc we obtain  
$L_\mathrm{111\, MHz}\sim 5.6\times 10^{26}$~erg~s${^{-1}}$. The corresponding  energy release is 
$E=\Delta t \times L_\mathrm{111\, MHz}\approx 7.2\times 10^{25}$~erg for $\Delta t=0.13$~s. 

The total amount of clean data used for the search for signals from J1527+3536 is about 210 hours. This indicates that low-frequency pulses from this source are quite rare. 

\section{Discussion}
\label{disc}

In this section, we briefly discuss several possibilities for the interpretation of the detected signal from J1527+3536. However, a detailed analysis of each of them is beyond the scope of this study, mostly because the existing data are insufficient to scrutinize the possibilities.  

\subsection{A source in the Galactic halo}

Formally, the DM value of the burst derived from the observations places the source in the Galactic halo. Thus, we begin the Discussion section with the possibility that the source of the burst is located within the halo. In this case, the source is not associated with the J1527+3536 system and may be related to another type of Galactic transient source. LPA already detected transient radio emitters at high Galactic latitudes. Below in this section, we briefly summarize these data and discuss their relation to the newly detected burst.

Almost 100 new transients with LPA3 have been discovered in a blind search\footnote{https://bsa-analytics.prao.ru/transients/rrat/}. The typical pulse widths (several tens of ms) coincide with the pulse widths of slow pulsars. The peak flux densities of the pulses are similar to the peak flux densities of known pulsars \citep{Tyulbashev2018}. The small observed dispersion measures (DM $ < 100$~pc cm$^{-3}$) indicate the Galactic nature of the sources. In the original papers, all the objects found were classified as Rotating RAdio Transients (RRATs). However, strictly speaking, we can call RRATs only transients that have a known period. So far, among the sources detected at LPA as radio transients, for just 19\% the periodicity is identified.

Among 81\% of the remaining transients, there are those for which only one pulse has been detected in the 3-year observation interval \citep{Tyulbashev2023}. There is a possibility that the pulse we found in this study is of the same nature as those previously discovered transients. At the same time, we cannot rule out the possibility that we have found a new type of transient.

Detailed estimation of the expected rate of detection of new transient bursts is not provided in previous studies related to the search for dispersed signals with LPA.\footnote{https://bsa-analytics.prao.ru/en/publications/}
However, it is possible to make a rough estimate for LPA3 based on general considerations. In the studies by \cite{Tyulbashev2018, Tyulbashev2018a, Tyulbashev2024a}, a search for transients was conducted in the same area ($-9^\circ < \delta < +42^\circ$; 16,500 sq.deg.) at time intervals of 1 month, 6 months, and 36 months. Correspondingly, 7, 34, and 91 transients were detected. That is, seven transients were discovered in the first month of observations. Then, (34-7)/5=5.4 transients per month were identified in the next 5 months. Finally, the rate (91-34)/30=1.9 transients per month was reported for the next 30 months. It can be seen that the number of transients discovered in one month is sharply decreasing, which is likely due to the finite number of transients in the Galaxy available to LPA3 in terms of sensitivity. 

Searching for pulsed radiation from J1527+3536, we processed 132 months of observations. Therefore, with a blind search in the range $-9^\circ < \delta < +42^\circ$, we naively expect to find no more than $(132-36) \times 1.9 \approx 180$ new transients. Since the size of the LPA3 beam pattern is $0.5^\circ \times 1^\circ$, we expect one new transient per (16500/0.5)/180=183 beams. That is, the probability of a random coincidence of the coordinates of the pulse detected by us with the coordinates of J1527+3536 is less than 1\%. We believe that this is a rather low probability for a chance coincidence. 

If some of the transients detected once belong to a new sample of sources, then using the same reasoning as shown above, we will get even more stringent estimates for the probability of a random coincidence of coordinates.


Still, if the source is in the Galactic halo, the isotropic energy release at the frequency 111 MHz is $\sim 10^{29}$--$10^{30}$~erg. This is not far from the values measured for RRATs \citep{2021A&A...647A.191B}. Thus, we provide an additional estimate of the probability of detecting a rotating radio transient in the Galactic halo, which is based on population considerations.
The birth rate of NSs high above the Galactic plane from runaway progenitors was recently calculated by \cite{2025arXiv250613660D}. Runaway stars constitute about 10\% of all core-collapse supernova (SN) progenitors. Thus, the rate of SN from runaway progenitors is $\sim 1/500$ per year. However, only 5\% of them happen at heights $|z|>1.9$~kpc. The direction towards the burst corresponds to the Galactic latitude $\approx 56^\circ$. I.e., the height above the Galactic plane is also about a few kpc. Thus, at a height corresponding to the hypothetical source of the burst the NS formation rate is $<1/10000$ per year. Typical characteristic ages of RRATs are $\sim 10^7$~yrs. Then in the whole Galaxy, we expect at most a few hundred of NSs with ages $\sim 10^7$~yrs at heights $\sim$~a few kpc.\footnote{Here we neglect a small number of sources which can reach such heights within 10 Myrs due to large kick velocities.} Not every NS is expected to demonstrate the RRAT-type activity. Then, we expect that the density of RRATs at high Galactic latitudes and at large heights above the Galactic plane is at most one per a few hundred square degrees. Thus, being conservative, we conclude that the probability of detecting a RRAT in the Galactic halo in a given direction by coincidence is negligibly small.  

\subsection{An extragalactic source} 

 As the measured value of the DM roughly corresponds to the DM through the whole Galactic halo in the direction of the burst, it is possible to speculate that the source is an extragalactic object at a distance $\sim$~a few Mpc where the DM still does not reach high values (see \citet{2025arXiv250412539T} for a review of the properties of the intergalactic medium). 

\begin{figure}[t]
    \centering  \includegraphics[width=1\linewidth]{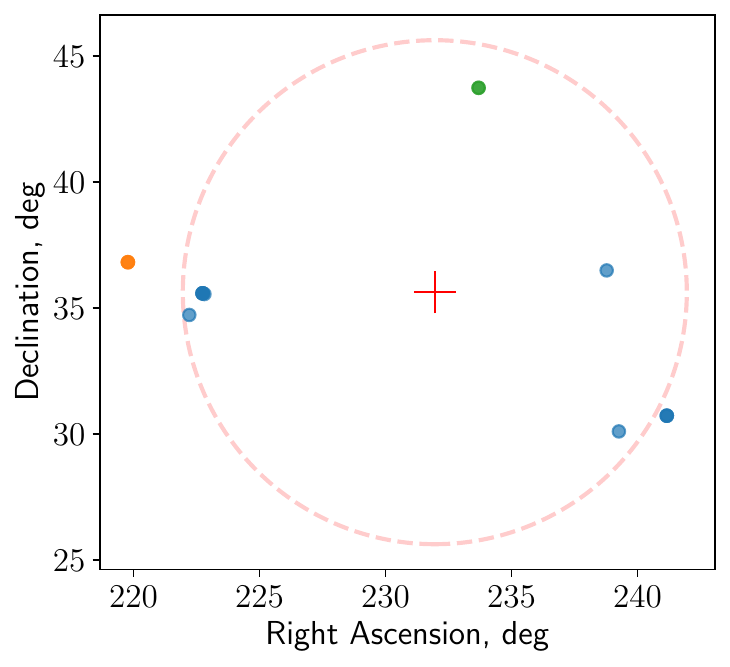}
    \caption{Objects near J1527+3536 in the sky. The position of J1527+3536 is marked with a red cross (15h27m48.48s +35d36m57.2s). The red circle with a 10-degree radius indicates the approximate area of the search. The dots represent objects found within this area: orange --- active galactic nuclei \citep{2019MNRAS.487.1823L}, blue --- galaxies \citep{2012ApJS..198....3W, 2022ApJS..261...21Y}, and green --- the Milky Way satellite low surface brightness galaxy Bootes IV \citep{2019PASJ...71...94H}.
    }
    \label{fig:gal} 
\end{figure}

Using the \verb|Astroquery| Python package \citep{2019AJ....157...98G}, we performed a search in {SIMBAD} database \citep{2000A&AS..143....9W} for galaxy-type objects in a 10-degree circle around the coordinates of the source J1527+3536 at a distance $<24$~Mpc.
The result is presented in Fig.~\ref{fig:gal}. Visibly, there are no good candidates, since the errors in the measured coordinates of the source are much smaller than $1^{\circ}$. We conclude that due to a small DM and the absence of galaxies close to the position of the burst in the sky, the source of the burst cannot be an extragalactic object.

\subsection{Additional dispersion in the system J1527+3536}

 As we demonstrated in the previous subsections, it is not easy to attribute the burst to a source in the Galactic halo or to an extragalactic object. Thus, despite the fact that the observed DM is too high for an object at a distance of 118 pc, it is reasonable to analyze various hypotheses relating the burst to the binary J1527+3536. However, in this case, it is necessary to discuss, at least superficially, the origin of the additional dispersion of the radio signal.

As the source is nearby and at a high Galactic latitude, it is unlikely that the additional DM is due to interstellar matter along the line of sight. Thus, we have to look at the possibilities to place the dispersing matter close to (or into) the binary system.

We cannot provide a clear explanation for an additional DM to bring the measure value into correspondence with the expectations. Still, some simple estimates can be made. One can consider the possibility that the DM is due to a cloud of dense material within the system J1527+3536. It is possible to derive some constraints on the properties of such a cloud.

The DM provides an estimate for the product of the electron density $n_\mathrm{e}$ and the size of the cloud $R_\mathrm{c}$. It is necessary to have the plasma frequency in the cloud $\nu_\mathrm{e}$ to be below the frequency of the detected signal $\sim 10^8$~Hz. 

\begin{equation}
    \nu_\mathrm{e} = \sqrt{\frac{n_\mathrm{e} e^2}{\pi m_\mathrm{e}}},
\end{equation}
where $e$ is an elementary charge and $m_\mathrm{e}$ is the electron mass.
Then, a cloud with radius $\sim 4\, R_\odot$ and mass $\sim 2 \times 10^{19}$~g can produce the necessary DM, still being transparent to low-frequency radio emission. Potentially, such a cloud can be produced by a huge coronal mass ejection, as observed flares of low-mass stars can be $\sim 10,000$ times more powerful than known solar flares \citep{2016ApJ...832..174O} and average solar coronal mass ejections have masses above $10^{15}$~g \citep{2019SSRv..215...39L}.  

Another possibility is related to the existence of a long-lived structure in the system. The stellar wind of a main sequence star with $M\approx 0.6\, M_\odot$ is $\sim 3\times10^{12}-2\times10^{13}$~g~s$^{-1}$ depending on the age of the star, as considered in detail in Sec.~\ref{discussion_NS}. This flow of matter can result in the formation of a gaseous structure responsible for the dispersion of the radio signal, since the mass $\sim2\times10^{19}$~g can be produced only in $\sim 10-100$ days. 
It is also possible that part of the stellar wind matter is captured by the WD and accreted, maybe even forming a disc \citep{2023ApJ...944L...4L}. Potentially, the accretion flow can produce the necessary effect of dispersion.


In the following subsections, we discuss several possibilities to explain the nature of the burst, assuming that it is related to the source J1527+3536.

\subsection{Stellar flares}

At first, we summarize arguments against the possibility that the detected signal can be due to a stellar flare (see a recent review on this type of transients in \cite{2024LRSP...21....1K}). 

The low-mass companion in J1527+3536 is classified as a late K dwarf \citep{2024ApJ...961L..48Z} or as a K9-M0 dwarf \citep{2023ApJ...944L...4L}. 
As it is demonstrated by \cite{2024A&A...684A...3Y}, see their Fig.~1, 
late-K/early-M dwarfs are more likely to flare in comparison with mid-M and early- to mid-K dwarfs.
Some of the stellar flares can be relatively short. E.g., 
Solar III type bursts observed at 150 MHz can have durations $\sim1$-3~s \citep{2013ApJ...762...60S, 2020A&A...639L...7V}.
Stellar flares can be as short as $\lesssim10$~ms \citep{1990SoPh..130..265B}. Thus, they have even been proposed as an explanation for FRBs by \cite{2014MNRAS.439L..46L}. However, this very short duration is not typical for low frequencies $\sim 10^8$~Hz. 
Also, the peak intensity from a source at $\sim 100$~pc can hardly reach $\sim   1$~Jy  \citep{2020A&A...639L...7V}.

Thus, we conclude that the pulse detected from J1527+3536 might not be a flare of the low-mass companion. 

\subsection{Neutron star as the source of the radio burst}
\label{discussion_NS}

 In the discovery paper, the authors advocated the hypothesis that the compact object in the binary system J1527+3536 is an NS \citep{2023ApJ...944L...4L}.\footnote{Note, that the proposal by \cite{2023ApJ...944L...4L} that the compact object is an NS similar to those of the Magnificent Seven \citep{2023Univ....9..273P} is not very probable, as typical ages of such NSs visible due to their thermal emission is rather short $\lesssim 10^6$~yrs, which is much less than a typical lifetime of a low-mass star.} 
 In addition, the short duration of the pulse --- $0.13$~s, --- indirectly points towards an NS as the source. 
 That is why we discuss several issues related to that hypothesis.

The observed properties of an NS are significantly dependent on the characteristics of the surrounding matter, which can be described by the velocity $v$ of the NS relative to the surrounding medium and the accretion rate $\dot{M}$.


In the case of a binary system, the velocity $v\approx\sqrt{v_\text{orb}^2+v_\text{w}^2+c_\text{s}^2}$, i.e., it is a combination of the relative orbital velocity $v_\text{orb}=\sqrt{G(M_*+M_\text{NS})/a}$, the wind velocity $v_\text{w}$, and the speed of sound $c_\text{s}$. 
With the known orbital period of the system J1527+3536, $0.2557$~days, and the masses of the two components, -- $M_*=0.62\,M_\odot$ and $M_\text{NS}=0.98\,M_\odot$ \citep{2024A&A...686A.299S}, -- we estimate the semi-major axis of a circular orbit $a\approx2R_\odot$ and $v_\text{orb}\approx400$~\kms. The radius of the main-sequence star $R_* \approx 0.6\,R_\odot$ \citep{2021A&ARv..29....4S}, so the NS is only $3.3 R_*$ away from the center of the second component. The profiles of the radial velocity of the stellar wind, modeled by \cite{2018PASJ...70...34S}, suggest that at $3.3R_*$ the wind accelerates to $\lesssim100$~\kms and has a temperature $\lesssim10^6$~K, which corresponds to $c_\text{s}\sim100$~\kms. So, the velocity $v=\sqrt{v_\text{orb}^2+v_\text{w}^2+c_\text{s}^2}\approx v_\text{orb}\approx400$~\kms.


In the binary system considered, the critical Roche lobe radius \citep{1983ApJ...268..368E} for the main-sequence star is $0.66\,R_\odot$, which is close to the star radius $\sim0.6\,R_\odot$, as noted by \cite{2024A&A...686A.299S}. The area of gravitational influence of the NS can be characterized by the gravitational capture radius $R_\text{G}=2GM_\text{NS}/v^2$. Given that $R_{\text{G}}\approx 1.2\,a \gtrsim a$ and the main-sequence companion nearly fills its Roche lobe, we assume that all wind particles emitted by the donor at a rate of $\dot{M}_w$ are captured by the NS, so we adopt $\dot{M} = \dot{M}_w$. 

Now, let us estimate the mass loss rate of the second component $\dot{M}_\text{w}$. For low-mass main-sequence stars, the following relation can be used: $\dot{M}_\text{w} \propto R_*^2M_*^{-3.36}\Omega_*^{1.33}$,
where $\Omega_*$ is the rotational frequency of the star. 
On timescales from several hundred Myr up to several Gyr, for the main sequence companion in J1527+3536, we can apply the general law of the spin frequency evolution proposed for the Sun \citep{2020ApJ...889..108A}. 
Given that $R_*$ is almost proportional to the mass of the star if $M_* \lesssim 1\,M_\odot$ \citep{2021A&ARv..29....4S}, $\dot{M}_\text{w}\propto M_*^{-1.36}$. With the known mass-loss rate of the Sun $\dot{M}_{\text{w}\odot} = 2\times10^{-14}\,M_\odot$~yr$^{-1}$ \citep{2021LRSP...18....3V}, the mass-loss rate of the star with $M_*=0.62\,M_\odot$ at the age of the Sun is 
\begin{equation}
\dot{M}_\text{w}(t=4.6\text{~Gyr})\approx4\times10^{-14}\,M_\odot\text{~yr}^{-1}\approx3\times10^{12}\,\text{g~s}^{-1}.
\end{equation}
According to \cite{2015A&A...577A..28J}, the mass loss rate varies with age as $\dot{M}_\text{w}\propto t^{-0.75}$ for $t>300$~Myr and remains constant for $t\le300$~Myr. So, the maximum mass-loss rate for the secondary component in the J1527+3536 system is 
\begin{equation}
\dot{M}_\text{w}(t\le0.3\text{~Gyr})\approx3\times10^{-13}\,M_\odot\text{~yr}^{-1}\approx2\times10^{13}\,\text{g~s}^{-1}.
\end{equation}
 Note that the significant uncertainty in the wind rate softens our assumption that $\dot{M} = \dot{M}_w$. 

We neglect the dependence of $\dot{M}_\text{w}$ on metallicity, since for stars with $M_*=0.6\,M_\odot$ the difference in $\dot{M}_\text{w}$ for the solar and zero metallicity is only $1.5-2$ times \citep{2018PASJ...70...34S}. 

\begin{figure}[t]
    \centering  \includegraphics[width=1\linewidth]{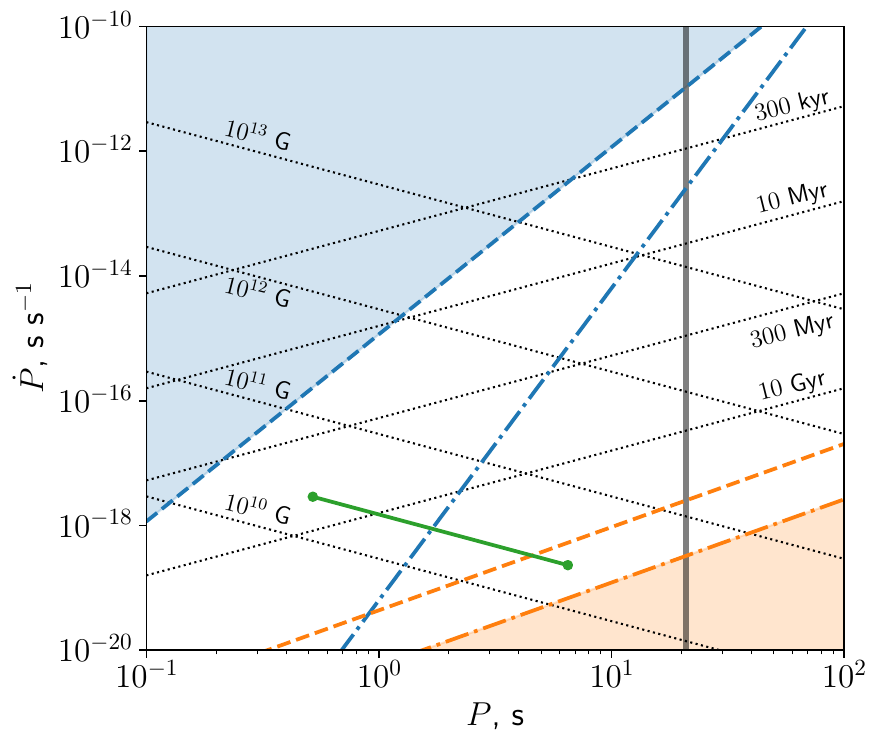}
    \caption{The period-period derivative diagram with diagonal lines for constant magnetic field values and constant ages of NSs at the ejector stage. The value of the magnetic field and spin period are applicable for all stages. The blue lines show the condition of termination of the ejector stage $R_\text{Sh}=R_\text{l}$. The orange lines are plotted for the condition of the onset of accretion $R_\text{A}=R_\text{cb}$. In both cases, the dashed line corresponds to the accretion rate $\dot{M}=3\times10^{12}$~g~s$^{-1}$, which is equal to $\dot{M}_\text{w}$ of a young secondary component ($\le300$~Myr), the dash-dotted line is for $\dot{M}=2\times10^{13}$~g~s$^{-1}$ ($\dot{M}_\mathrm{w}$ of the star with the age $4.6$~Gyr). The vertical gray line is the maximum $P$, which is derived from the condition $R_\text{l}=a-R_*$. The blue area shows the parameters for an NS at the ejector stage if the secondary component is young ($\le300$~Myr). The orange area is for an accreting NS with an older secondary component ($4.6$~Gyr). The green segment represents the parameters of the NS derived by \cite{2023ApJ...944L...4L}: the spin period $0.52\text{~s} \le P \le 6.50\text{~s}$, the magnetic field value 
    $B= 2.27\times10^{10}$~G.
    }
    \label{fig_p_p_dot}
\end{figure}

Having determined the characteristics of the medium surrounding the NS ($v=400$~\kms, $\dot{M}=\dot{M}_\text{w}$), we can now explore the regimes of interaction of the NS with the external material. In Fig.~\ref{fig_p_p_dot}, we present the potential characteristics of the NS in different regimes. Note, that the values of $\dot P$ and characteristic ages, given in the figure, are relevant only for the ejector stage. However, the value of the magnetic field and spin period are applicable for all stages.


There is no evidence of an extensive interaction between the magnetosphere of the NS and the secondary component. Therefore, the radius of the magnetosphere should be less than the distance between the compact object and the surface of its companion. The maximum radius of the magnetosphere is the light cylinder radius $R_\text{l}=c/\omega$, where $\omega=2\pi/P$ is the spin frequency and $c$ is the speed of light. So, the maximum spin period of the NS in the J1527+3536 system, derived from the condition $R_\text{l}=a-R_*$, is approximately $20$~s.


\subsubsection{Ejector stage}

First, we assume that the NS is an ejecting pulsar and describe it following the general approach considered by \cite{1992ans..book.....L, 2024Galax..12....7A}. During the ejector stage, the NS loses its rotational energy at a rate $L=2\mu^2\omega^4/c^3$ \citep[see e.g.,][]{2024PASA...41...14A}, where $\mu=BR_\text{NS}^3$ is the magnetic dipole moment ($R_\text{NS}=10$~km). The spin period derivative can be expressed as
\begin{equation}
\label{P_dot}
    \dot{P} = \frac{8\pi^2}{Ic^3} \frac{\mu^2}{P} \approx 3\times10^{-15} \left(\frac{P}{1\text{~s}}\right)^{-1} \left(\frac{B}{10^{12}\text{~G}}\right)^2 \, \text{s~s}^{-1} ,
\end{equation}
where $I=10^{45}$~g~s$^{-1}$ is the moment of inertia of an NS. 

The characteristic radius of the interaction of the pulsar wind with the external matter can be estimated as the distance from the NS, where the pressure of the pulsar wind $p_\text{in} = L/(4\pi R^2 c)$ is equal to the ram pressure $p_\text{out} = \rho v^2$. Here, $\rho$ is the density of the external material, which can be derived from the continuity equation as $\rho = \dot{M}/(4\pi R_\text{G}^2 v)$. The resulting distance from the NS is known as the Shvartsman radius:
\begin{equation}
    \label{R_Sh}
    R_{\text{Sh}}=\left( \frac{8 \mu^2 (GM_\text{NS})^2 \omega^4}{\dot{M} v^5 c^4} \right)^{1/2}.
\end{equation}
As long as $R_\text{Sh}$ exceeds the light cylinder radius, the material cannot enter the magnetosphere of the NS, so the NS remains at the ejector stage.

Typically, NSs at this stage are young objects. Thus, we use $\dot{M}=\dot{M}_\text{w}(t\le300\text{~Myr})$ to calculate the parameters of an NS at the ejector stage. This is the blue area in Fig.~\ref{fig_p_p_dot}.

Two mechanisms of a radio burst from an NS at the ejector stage in a binary system are presented and discussed by \cite{1984Ap&SS..98..221L, 2008arXiv0812.4587P}. 
According to the first hypothesis, if the radio pulsar is surrounded by material from the wind, a cavern with radius $R_\text{Sh}$ can form around the NS. As the cavern is filled with the pulsar's wind, the internal pressure increases, potentially causing the cavity to open and resulting in a radio burst. These are periodic events with a characteristic time interval:
\begin{equation}
    t_\text{br} = \frac{R_\text{Sh}}{c}\left(\sqrt{\frac{2k_\text{B}T_\text{w}}{m_\text{p}v^2}} - \frac{R_\text{Sh}}{a}\right)^{-1},
\end{equation}
where $k_\text{B}$ is the Boltzmann constant, $m_\text{p}$ is the proton mass and $T_\text{w}$ is the temperature of the wind. Given that $T_\text{w}\sim10^6$~K \citep{2018PASJ...70...34S}, the ratio $\sqrt{{2k_\text{B}T_\text{w}}/({m_\text{p}v^2})} \approx0.3$, so the time interval between radio bursts $t_\text{br}$ can be arbitrarily large if $R_\text{Sh}$ is very close to $0.3\,a$. Otherwise, if $R_\text{Sh}/a\lesssim 0.2$, then $t_\text{br}\lesssim 10R_\text{Sh}/c\lesssim 10a/c \sim 50$~s.

In the other case, after the opening of the cavern, the envelope filled with the pulsar wind is considered to be detached from the NS. The interval between bursts is $t_\text{b} = (a/R_\text{Sh})^2 R_\text{G}/c$. As in the previous case, for $R_\text{Sh}\sim R_\text{G} \sim a$, the periodicity of radio bursts is of the order of minutes.

The two scenarios are similar in terms of burst duration and the energy released. When the cavity collapses, it produces a radio burst with a duration of $\Delta t_\text{b} = R_\text{G}/c\approx 8\text{~s}\,(v/400\text{~\kms})^{-2}$. The released energy can be estimated as $E\sim L\Delta t \approx 6\times10^{33}\text{~erg~s}^{-1}\,(P/1\text{~s})^{-4}(B/10^{12}\text{~G})^2(\Delta t/50\text{~s})$.

In both cases, the energy released is sufficient to produce a burst with the observed $L_\text{radio}$. However, the predicted burst duration is longer than the observed radio signal, and the time interval between bursts is relatively short -- of the order of minutes. The lack of observed repetition in the radio source suggests that the period of the bursts may be long, making these interpretations unlikely.

\subsubsection{Propeller and georotator stages}

If the age of the NS is about several billion years, the NS can be at the propeller stage. 
This stage is considered by \cite{2023ApJ...944L...4L} to provide interpretation of an X-ray emission, which is assumed to be generated through the accretion process onto the magnetosphere. The radius of the magnetosphere is the Alfv{\'e}n radius: 
\begin{equation}
    \label{R_A}
    R_A = \left( \frac{ \mu^2}{2 \dot{M} \sqrt{2GM}}  \right)^{2/7}.
\end{equation}

The NS stays at the propeller stage as long as the centrifugal barrier prevents the captured material from falling onto the surface of the compact object. The condition for the onset of accretion can be written as $R_\text{A} \le R_\text{cb}$, where $R_\text{cb}=0.87(GM/\omega^2)^{1/3}$ \citep{2023MNRAS.520.4315L}.

According to \cite{2023ApJ...944L...4L}, the X-ray emission of the object is too weak for the NS to be at the accretor stage, so we do not consider possible mechanisms of radio emission of the NS with the parameters of an accretor. In Fig.~\ref{fig_p_p_dot}, the corresponding region with parameters $P$ and $B$ of an NS at this stage is shown in orange. The parameters are calculated taking into account that the age of the companion is several billion years (we use $4.6$~Gyr). To display the region that corresponds to accretors in the $P-\dot{P}$ diagram, eq.~\ref{P_dot} is used. The rest of the diagram is white. This part corresponds to the propeller stage.

We discuss some exotic mechanisms for transient radio emission by a mature/old NS, which does not accrete the material onto its surface in a continuous way, as outlined by \cite{2008arXiv0812.4587P}. For example, if the magnetospheric boundary of the NS at the propeller stage can extend to the light cylinder radius, a region of open magnetic field lines may form, leading to the generation of radio emission. In this scenario, the burst frequency would correspond to the spin period of the NS. As mentioned above, the rotation period of the NS in the system cannot exceed $\approx20$~s, but radio observations do not reveal any periodicity with a time interval comparable to $20$~s (or repetition on the scale of years).


Another late evolutionary stage is the georotator stage. When the gravitational influence of the neutron star at the magnetospheric boundary becomes negligible, i.e. the magnetosphere radius $R_\text{A}$ exceeds $R_\text{G}$, the material will not accrete onto the surface even if the condition for the propeller-accretor transition ($R_\text{A}\le R_\text{cb}$) is satisfied. This is known as the georotator stage. As considered by \cite{2001AIPC..586..519R, 2001ApJ...561..964T}, during this phase, reconnection of magnetic field lines in the tail of the magnetosphere can occur. For the high density of the surrounding medium $n\gtrsim1$~cm$^{-3}$, the rate of energy released is
\begin{equation}
    \dot{E}_\text{rec} = 5\times10^{21}  \left(\frac{B}{10^{12}\text{~G}}\right)^{2/3} \left(\frac{n}{1\text{~cm}^{-3}}\right)^{2/3} \left(\frac{v}{400\text{~\kms}}\right)^{7/3}\text{~erg~s}^{-1}.
\end{equation}
Given that the velocity of the wind at the position of the NS is $v_\text{w}\sim100$~\kms, the number density $n$ can be derived from the continuity equation $n=\dot{M}_\text{w}/(4\pi a^2 v_\text{w}m_\text{p})$. Using this, the possible energy release in the system is estimated as $\dot{E}_\text{rec} = 0.4-1.4\times10^{26}\,(B/10^{12}\text{~G})\, \text{~erg~s}^{-1}$, which is sufficient for the observed $L_\text{radio}$ if the magnetic field is high enough. However, the NS in the J1527+3536 system cannot be at the georotator stage. This stage requires that the magnetosphere radius exceeds the gravitational capture radius. The condition $R_\text{A}>R_\text{G}$ is satisfied only if the magnetic field of the NS is high -- $B\gtrsim3\times10^{14}$~G. This is not very probable for an older NS, and no magnetar-like activity is observed from the source. In addition, the condition $R_\text{A}\le R_\text{cb}$ should also be met, which places the NS in the orange area of Fig.~\ref{fig_p_p_dot}, where only the NSs with $B\lesssim10^{11}$~G are present. So, this scenario is also unlikely.

Finally, we consider the transient propeller stage \citep{2008arXiv0812.4587P}. It is characterized by periodic collapses of an envelope formed around the magnetosphere of the NS at the propeller stage. As a result of the collapse, an X-ray and radio burst can be expected. The duration of the burst is $\Delta t_\text{p} \sim R_\text{cb}/ v_\text{ff} \sim \sqrt{R_\text{cb}^3/(2GM_\text{NS})}\sim 0.08\text{~s}\,(P/1\text{~s})$. If $\Delta t = 0.13$~s, then the spin period of the NS is $P\approx1.7$~s. In the regime considered, the spin period should be long enough for the NS to be at the propeller stage instead of the ejector stage. From the Fig.~\ref{fig_p_p_dot} we can conclude that the magnetic field of the NS with $P=1.7$~s at the propeller stage is $B<10^{12}$~G if the system is young ($\lesssim300$~Myr), and $B\lesssim3\times10^{10}$~G if the second component is several billion years old. Among all scenarios considered, the transient propeller stage appears to be the most plausible explanation of the radio burst, if the primary component of the J1527+3536 binary system is an NS.

\subsection{Radio signals from white dwarf binaries}

Already, the first mass estimate of the compact object made by \cite{2023ApJ...944L...4L} has provided a relatively low value $0.98\pm 0.03 \, M_\odot$. Generally, this value is more consistent with a WD than with an NS.
More recently, \cite{2024ApJ...961L..48Z} 
presented new observations and detailed analysis of the system J1527+3536.
In particular, they obtained a new mass estimate for the compact object in the system. With the value $0.69\pm 0.02\, M_\odot$,  it can hardly be an NS   \citep{2024arXiv240711153C}.\footnote{See, however, discussion about the sources HESS J1731-347 \citep{2023ApJ...958...49S} and PSR J1231‑1411 \citep{2024ApJ...976...58S} for which low-masses have been reported on the basis of indirect model-dependent measurements. } 
 This result is based on high-resolution spectroscopy, which allowed for an independent determination of the orbital inclination. The obtained value ($60^\circ-65^\circ$) is greater than the previous result by \cite{2023ApJ...944L...4L} who estimated this parameter as $\approx 45^\circ$. 
The system is also detected in X-rays and its luminosity can be explained by an accreting WD \citep{2023ApJ...944L...4L, 2024A&A...686A.299S}.
Thus, a WD companion seems to be a better justified hypothesis according to the present-day data. 

 Among WD binaries with low-mass companions and short orbital periods, there are several known radio emitters (see e.g., \cite{2024MNRAS.531.1805P, 2025A&A...695L...8R} and references therein). They demonstrate different types of activity, including transient radio bursts, which are, in some respects, similar to the one we detected from the direction of J1527+3536. Thus, we consider it a valuable possibility that J1527+3536 is a WD~--~red dwarf system with sporadic activity at radio wavelengths. 
In the following, we describe several WD binaries known to be pulsating radio sources and a few similar systems. Their properties are presented in the Table~3 together with J1527+3536. 




Recently, a group of so-called long-period radio transients (LPRTs) has been identified. 
At the moment, it includes about 10 sources. They are characterized with transient radio emission with periodicity from several minutes up to several hours. In two cases (GLEAM-X J0704-37, ILT J1101 + 5521), the sources are identified as low-mass binary systems with WDs. 
As it is visible, in some respects, they are alike J1527+3536, having slightly shorter orbital periods and lower masses of normal stars, but similar radio luminosities.  

GLEAM-X J0704-37 was detected by MWA \citep{2024ApJ...976L..21H}.
The source has a very steep radio spectrum with $\alpha=6.2\pm0.6$. An optical counterpart is consistent with an M5 dwarf. The binary period (2.9 hours) was recently established by \cite{2025A&A...695L...8R}. This author also estimates the distance as $\approx 400$~pc; therefore, the radio luminosity can be slightly lower than the value provided by \cite{2024ApJ...976L..21H} and given in the table. 

ILT J1101+5521 was detected by LOFAR \citep{2025NatAs...9..672D}.
This source also has a steep spectrum with $\alpha=4.1\pm1.1$. For the distance 504 pc the isotropic radio luminosity during pulses is $\sim10^{27}$-$10^{28}$~erg~s$^{-1}$. 

The next two sources are so-called WD pulsars. For them, it is established that the pulsating period is related (directly or via the beat frequency) to the spin period of the WD, and the emitted power is provided by the spin-down of the compact object. 

 AR Sco was discovered by \cite{2016Natur.537..374M}. 
The distance to the source is $\approx 116$~pc. The mass of the stellar companion is in the range 0.2-$0.45\, M_\odot$, probably closer to the lower end.  The observed periodicity corresponds to the beat frequency between the spin (1.95 min) and the orbital period (2.9 hours). The maximum total luminosity is $6.3 \times 10^{32}$~erg~s$^{-1}$. The source was detected in radio by ATCA (9 GHz) with the flux $\sim12-16$~mJy, which roughly corresponds to the isotropic luminosity $\sim 2\times 10^{27}$~erg~s$^{-1}$. The radio pulses are very wide (about 50\% of the cycle). The source is also detected at higher frequencies (submillimeters) \citep{2025ApJ...986...78B}. These submillimeter observations at 200-400 GHz provided fluxes $\sim 0.1$~Jy which corresponds to the luminosity $\sim$few$\times10^{29}$~erg~s$^{-1}$. 

For eRASSU J1912-4410, the maximum radio luminosity during flares can be estimated from the distance 237 pc and flux obtained during MeerKAT observations in the L-band (1.4 GHz) equal to $\approx 10$~mJy \citep{2023NatAs...7..931P}. Then, we have $L_\mathrm{radio}\approx 10^{27}$~erg~s$^{-1}$. The periodicity is due to the spin period. 

Next, we present two radio-emitting WD binaries that can contain a WD at the propeller stage \citep{1997MNRAS.286..436W}. 
These are long-studied systems, AE Aqr and its recently discovered twin, LAMOST J024048.51+195226.9. Both systems have radio luminosities, companion masses, and orbital periods similar to those of J1527+3536.

Radio emission from the WD binary AE Aqr was detected long ago \citep{1987ApJ...323L.131B}. 
Recently, radio emission pulsating at the spin period 33 s was also reported by \cite{2022heas.confE..46M} 
who observed with MeerKAT. At $\sim 1.1$~GHz the flux reached $\sim8$~mJy. 
At a distance of 90 pc, this corresponds to the luminosity $\sim 8\times 10^{25}$~erg~s$^{-1}$. The pulses are wide, comparable to the pulsating period.
AE Aqr is also a gamma-ray source \citep{2022heas.confE..46M}.

LAMOST J024048.51+195226.9 was recently confirmed as a twin of AE Aqr \citep{2021ApJ...917...22G, 2021MNRAS.503.3692P}.
Radio observations were done by MeerKAT \citep{2021MNRAS.503.3692P}.
The distance to the source is $\approx 620$~pc. The flux at 1.5 GHz is $\sim 0.6$~mJy. Thus, the radio luminosity can be estimated as $\sim 2.5\times 10^{26}$~erg~s$^{-1}$. There was just one observation in radio. The source demonstrated variability on the scale from $\sim 10$~s to minutes, but no periodicity was detected.  As the companion is classified as an M1.5 dwarf, its mass is $\sim 0.5 \, M_\odot$.

The last source in the table -- SDSS J230641.47+244055.8 -- is not observed as a radio pulsating object. It was proposed as a system similar to AR Sco  \citep{2025arXiv250620455C}. 

\begin{table*}
    \begin{tabular}{lccccccc}
        \hline
        \\
    Object & $P_\mathrm{pulse}$, m &$P_\mathrm{spin}, $~s 
    & $P_\mathrm{orb}$, h & $\Delta t$, s & $L_\mathrm{radio}$, erg~s$^{-1}$ & $M_*, \, M_\odot$ &  Ref. \\

    \\
    \hline
    \\
J1527+3536 & --- & --- &  6.14 & 0.13 & $5.6\times 10^{26}$ & 0.62 &  \cite{2023ApJ...944L...4L} and this work\\
GLEAM-X J0704-37 & 175 & --- & 2.9 & $\sim30$ & $\sim10^{26}$& 0.14 &  \cite{2024ApJ...976L..21H}\\
ILT J1101 + 5521 & 125.5 & 7530 (?) & 2.1 & $\sim (30$-90)& $\sim10^{27}$-$10^{28}$ & $\approx 0.2$ & \cite{2025NatAs...9..672D}\\
AR Sco & 1.97  & 117 & 3.56 & $\sim 100$ & $\sim 10^{27}$& $\approx 0.3 $& \cite{2016Natur.537..374M} \\
J1912-4410 & 5.3 & 319  &  4.03 & $\lesssim 4$ & $\approx 10^{27} $& $\approx 0.3$ & \cite{2023NatAs...7..931P}\\
AE Aqr & 0.55 & 33 & 9.9 & $\sim 10$& $8\times10^{25}$& $\sim 0.6$& \cite{2022heas.confE..46M}\\
 J024048.51+195226.9 & --- & --- & 7.3 & --- & $\sim2.5\times 10^{26}$ & $\sim 0.5$ & \cite{2021MNRAS.503.3692P}\\
 J230641.47+244055.8& --- & 92 & 3.49 & --- & --- &  0.19-0.28&  \cite{2025arXiv250620455C}\\
    \hline
    
    \end{tabular}
    \caption{Radio emitting binaries with WDs and similar systems. Here, $M_*$ is the mass of a main sequence companion.}
\end{table*}

 As can be seen in Table 3, the properties of J1527+3536 resemble those of other radio-emitting WD binaries. However, the burst rate of J1527+3536  is very low, while the single detected pulse is very short compared to other sources. 

 The latter property can be explained if the spin period of the WD in J1527+3536 is very short. If this is the case, the system might have a very non-trivial evolution to allow for accretion-supported spin-up of the compact object at preceding evolutionary stages. 
 Potentially, if the WD magnetic field is not very high, it is easier to spin up the compact object by accretion \citep{1994PASP..106..209P}. For the spin period of about a few tens of seconds, even a field below 1~MG can easily explain the observed energetics. On the other hand, a low field can explain the absence of continuous interaction of the magnetosphere with the plasma from the K-dwarf, and so the absence of radio detection out of the flare episode. \footnote{
 \cite{2023ApJ...944L...4L} gave an upper limit on the radio flux from J1527+3536 as $5\mu$Jy at 1.2 GHz.}

 One can speculate that the burst of J1527+3536 can be related to a massive coronal mass ejection (CME) by the low-mass companion. The interaction of the plasma blob with a rapidly rotating magnetosphere of the WD could produce a radio flare. The combination of a strong CME towards the white dwarf and a beamed emission towards the Earth can be a rare occasion. This explains the low rate of observed events. Weaker pulses might be more frequent as weak CMEs happen more often. 
 In the LPA archival data, we found evidence in favor of several weak pulses in the direction of J1527+3536 with the same DM. However, the S/N ratio ($  \sim 5$) is not sufficient to claim the discovery \citep{Eldarov2025}.





\section{Conclusions}
\label{conc}

In this paper, we presented results of an archival search for radio emission from five recently discovered binary systems with dim (or invisible) compact objects, predominantly neutron stars. We used data obtained with the LPA at the Pushchino Radio Astronomy Observatory at a low frequency (111 MHz). 
For all the systems, we do not find any persistent or periodic radio sources. 

In one case, J1527+3536, we identified a single radio burst with duration $\approx  0.13$~s and flux density $\approx 13$~Jy. The compact object in this system can be a neutron star or a white dwarf, as its mass is not certain. We discussed several possibilities to interpret the detected burst. We conclude that this system can be similar to white dwarf binaries with detected radio emission (long-period radio transients, AE Aqr-like, and AR Sco-like systems). More observations are necessary to uncover the nature of J1527+3536. 


\section*{Acknowledgements}

MDA and SBP are supported by the RSF grant 25-12-00012. The authors thank S. Andrianov for technical support of the work. This research has made use of NASA's Astrophysics Data System Bibliographic Services.

\bibliographystyle{elsarticle-harv} 
\bibliography{wide_radio}

\end{document}